\documentclass[aip,jcp,reprint,amsmath,amssymb,longbibliography,lengthcheck,showpacs,superscriptaddress,floatfix]{revtex4-1}
\usepackage{enumerate}
\usepackage{graphicx}
\usepackage{color}
\begin{document}
%
\title{Phonon transport properties of particulate physical gels}
%
\author{Hideyuki Mizuno}
\email{hideyuki.mizuno@phys.c.u-tokyo.ac.jp}
\affiliation{Graduate School of Arts and Sciences, The University of Tokyo, Tokyo 153-8902, Japan}
\author{Makoto Hachiya}
\affiliation{Graduate School of Arts and Sciences, The University of Tokyo, Tokyo 153-8902, Japan}
\author{Atsushi Ikeda}
\email{atsushi.ikeda@phys.c.u-tokyo.ac.jp}
\affiliation{Graduate School of Arts and Sciences, The University of Tokyo, Tokyo 153-8902, Japan}
\affiliation{Research Center for Complex Systems Biology, Universal Biology Institute, The University of Tokyo, Tokyo 153-8902, Japan}
%
\date{\today}
%
\begin{abstract}
Particulate physical gels are sparse, low-density amorphous materials in which clusters of glasses are connected to form a heterogeneous network structure.
This structure is characterized by two length scales, $\xi_s$ and $\xi_G$: $\xi_s$ measures the length of heterogeneities in the network structure, and $\xi_G$ is the size of glassy clusters.
Accordingly, the vibrational states of such a material also exhibit a multiscale nature with two characteristic frequencies, $\omega_\ast$ and $\omega_G$, which are associated with $\xi_s$ and $\xi_G$, respectively: (i) phonon-like vibrations in the homogeneous medium at $\omega < \omega_\ast$, (ii) phonon-like vibrations in the heterogeneous medium at $\omega_\ast < \omega < \omega_G$, and (iii) disordered vibrations in the glassy clusters at $\omega > \omega_G$.
Here, we demonstrate that the multiscale characteristics seen in the static structures and vibrational states also extend to the phonon transport properties.
Phonon transport exhibits two distinct crossovers at the frequencies $\omega_\ast$ and $\omega_G$~(or at wavenumbers of $\sim \xi_s^{-1}$ and $\sim \xi_G^{-1}$).
In particular, both transverse and longitudinal phonons cross over between Rayleigh scattering at $\omega < \omega_\ast$ and diffusive damping at $\omega>\omega_\ast$.
Remarkably, the Ioffe--Regel limit is located at the very low frequency of $\omega_\ast$. Thus, phonon transport is localized above $\omega_\ast$, even where phonon-like vibrational states persist.
This markedly strong scattering behavior is caused by the sparse, porous structure of the gel.
\end{abstract}
%
\maketitle
%
\section{Introduction}~\label{sect:intro}
Amorphous materials are ubiquitous in our daily lives.
Glasses are known to be high-density amorphous materials~\cite{Phillips_1981,Berthier2011b,larson1999structure} in which particles are tightly packed to form a disordered but homogeneous configuration.
In contrast to glasses, gels are low-density amorphous materials~\cite{larson1999structure,de1979scaling,mewis2012colloidal}.
In particulate gels, the constituent particles form a sparse network structure or a porous structure.
Gelation of a particular system can be realized via several distinct methods~\cite{Zaccarelli2007,Sciortino_2011,Lu2013,Ruiz_2021}.
For example, we can introduce long-range repulsive forces and competing attractive forces between the constituent particles~\cite{Mossa_2004,Sciortino_2004,Campbell_2005,Toledano_2009}.
Such interactions generate stable clusters, which then form a network structure to realize a gel state.
Additionally, we can consider patchy particles that interact via limited-valency potentials~\cite{Bianchi_2006,Michele_2006,Zaccarelli_2006}.
Such a system maintains an equilibrium liquid state even at low densities without phase separation dynamics, and it then undergoes an arrested transition to the gel state as the temperature is lowered.

In addition, there is another gelation process, so-called \textit{arrested phase separation}~\cite{Zaccarelli2007,Lu2008,Zaccarelli_2008,Lu2013}.
A system composed of attractive particles, e.g., Lennard--Jones~(LJ) particles, generally undergoes phase separation between the low-density gas phase and the high-density liquid phase when the density is low enough~\cite{Yamamoto_1994,Foffi_2002,Bailey_2007}.
In this situation, if the system is cooled to low temperatures, this phase separation process is interrupted since the liquid phase undergoes the glass transition~\cite{Testard_2011,Testard_2014}.
Note that similar phase separation kinetics are also observed in a binary mixture of glass-forming liquids, in which liquid--liquid phase separation is interrupted due to the glass transition~\cite{oku2020phase}.
In the end, glassy clusters form in which the particles are densely packed, these clusters are dispersed in space, and they then become connected to form a sparse network structure.
In the present work, we focus on such nonequilibrium particulate gels (particulate physical gels).

In our recent work, Ref.~\onlinecite{Mizuno_2021}, we have provided a comprehensive understanding of the structural, mechanical, and vibrational properties of particulate physical gels.
In particular, we have elucidated the multiscale nature of their vibrational states, for which there are two characteristic frequencies $\omega_\ast$ and $\omega_G$, i.e., (i) phonon-like vibrations occur in the homogeneous media at low frequencies below $\omega_\ast$, (ii) phonon-like vibrations occur in the heterogeneous media with fractal structures at intermediate frequencies between $\omega_\ast$ and $\omega_G$, and (iii) disordered vibrations occur in the glassy clusters at high frequencies above $\omega_G$.
Accordingly, the vibrational density of states~(vDOS) $g(\omega)$ shows crossover behaviors at the frequencies $\omega_\ast$ and $\omega_G$, with a characteristic plateau appearing at $\omega_\ast < \omega < \omega_G$.
This behavior of the vDOS has also been observed in an early experiment on silica aerogels~\cite{Vacher_1990}.

As a complement to Ref.~\onlinecite{Mizuno_2021}, the present work focuses on the phonon transport properties of particulate gels.
In a previous series of experimental works~\cite{Courtens_1987,Courtens_1988,Vacher_1989,Anglaret_1994}, including the abovementioned work~\cite{Vacher_1990}, scattering experiments have been performed on silica aerogels, such as Brillouin, Raman, and inelastic neutron scattering, and the dispersion curves and line widths have been measured for acoustic excitations~(phonon transport).
These works have demonstrated that acoustic excitations show crossover behaviors that are directly associated with those in the vDOS: both the acoustic excitations and the vDOS exhibit crossovers at the same frequency points.
The experimental observations reported in these works have also been discussed in terms of so-called fractons, which are highly localized vibrations in the fractal structure~\cite{Alexander_1989,Yakubo_1989,percolation,Nakayama_1994}.

Here, we provide an understanding of phonon transport properties based on structural, mechanical, and vibrational properties~\cite{Mizuno_2021}.
We reveal that the multiscale characteristics persist in phonon transport, which exhibits two distinct crossovers at frequencies of $\omega_\ast$ and $\omega_G$, i.e., (i) Rayleigh scattering behavior at low frequencies below $\omega_\ast$, (ii) diffusive damping behavior at intermediate frequencies between $\omega_\ast$ and $\omega_G$, and (iii) phonon transport through glassy clusters at high frequencies above $\omega_G$.
We also find that the Ioffe--Regel~(IR) limit is located at the remarkably low frequency of $\omega_\ast$.
Phonons are therefore localized at frequencies above $\omega_\ast$, even though phonon-like vibrational states are maintained in this frequency regime.
We will discuss our results with reference to experimental observations~\cite{Vacher_1990,Courtens_1987,Courtens_1988,Vacher_1989,Anglaret_1994} and in terms of their relevance to fractons~\cite{Alexander_1989,Yakubo_1989,percolation,Nakayama_1994}.

\begin{figure}[t]
\centering
\includegraphics[width=0.495\textwidth]{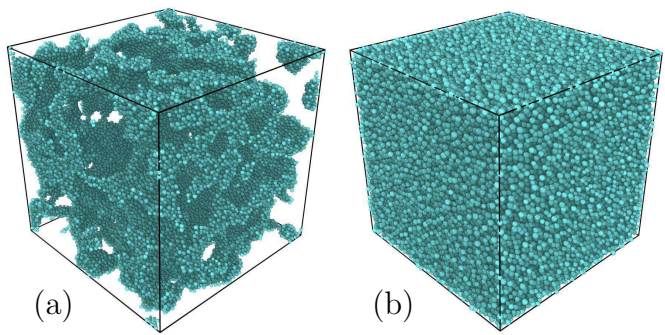}
\caption{\label{fig_visual}
{Particle configurations of gels and glasses.}
We visualize (a) a gel state with $\rho = 0.3$~($\varphi = 0.16$) and (b) a glass state with $\rho = 1.0$~($\varphi = 0.54$).
The number of particles is $N=80000$, and the system lengths are (a) $L=64.4$ and (b) $L=43.1$ in LJ units.
}
\end{figure}

\section{Summary of our previous paper~\cite{Mizuno_2021}}~\label{sec:preliminaries}
In this section, we provide a summary of our previous paper~\cite{Mizuno_2021}, which reported comprehensive results on the structural, mechanical, and vibrational properties of a model of particulate physical gels.
The model considers the zero-temperature quenching of a system composed of polydisperse LJ particles in three-dimensional~($d=3$) space~($d$ denotes the number of spatial dimensions).
The particles interact through the LJ potential given in Eq.~(\ref{eq:ljpotential}).
A control parameter of this model is the (number) density $\rho$, or the volume fraction $\varphi$~(defined in Eq.~(\ref{eq:volumef})).
We can realize gel states through arrested phase separation by setting relatively low values of $\rho$.
Note that the present system is specifically an ``LJ gel'' but is referred to simply as a ``gel'' throughout this paper.

\subsection{Structural properties}
In a gel of the type considered here, clusters of particles in which the particles are tightly packed in amorphous states are dispersed in space, and these clusters are connected to form a sparse network structure.
As visualized in Fig.~\ref{fig_visual}(a), we observe a porous structure with ``holes'' throughout the system, which is in contrast to the homogeneous structure characteristic of the dense packing of glasses illustrated in Fig.~\ref{fig_visual}(b).
In the gel structure, there are two characteristic lengths, $\xi_s$ and $\xi_G$: $\xi_s$ is the length scale of heterogeneities in the network structure, and $\xi_G$ is the size of the glassy clusters.
Thus, a gel possesses a multiscale structure: it acts as (i) a homogeneous medium at long lengths~($r > \xi_s$), (ii) a heterogeneous medium with a fractal structure at intermediate lengths~($\xi_s > r > \xi_G$), and (iii) an amorphously structured medium at short lengths~($r < \xi_G$).
As the density $\rho$ decreases, $\xi_s$ grows, following a power-law scaling of
\begin{equation}
\xi_s \propto \rho^{-0.7}, \label{eq.length}
\end{equation}
whereas $\xi_G$ is insensitive to $\rho$, and its corresponding wavenumber is approximately $q_G = 2\pi/\xi_G \simeq 0.8$~(i.e., $\xi_G \simeq 7.8$) in LJ units.

\subsection{Mechanical properties}
Due to their sparse structure~(porous structure), gels can be extremely soft, with elastic moduli orders of magnitude smaller than those of glasses.
Both the shear modulus $G$ and the bulk modulus $K$ significantly decrease with decreasing $\rho$, following power-law scaling behaviors of
\begin{equation}
G_0 \propto \rho^{2.8}, \quad G_\text{ave} \propto \rho^{2.5}, \quad K \propto \rho^{2.5}. \label{eq.moduli}
\end{equation}
Here, we note that gels show anisotropic shear elasticity, and the five components of the shear modulus have different values.
In Eq.~(\ref{eq.moduli}), $G_0$ denotes the lowest component, while $G_\text{ave}$ denotes the average value over the five components.

\subsection{Vibrational properties}
Due to the multiscale structure of a gel, there are two characteristic frequencies, $\omega_\ast$ and $\omega_G$, relevant to its vibrational properties, which are associated with the two lengths $\xi_s$ and $\xi_G$, respectively.
The associations between these frequencies~($\omega_\ast$ and $\omega_G$) and lengths~($\xi_s$ and $\xi_G$) are described in terms of appropriate sound speeds~\cite{Leonforte2005}, as follows:
\begin{equation}
\begin{aligned}
& \omega_\ast \propto \sqrt{\frac{G_0}{\rho}} q_s = \sqrt{\frac{G_0}{\rho}} \frac{2\pi}{\xi_s} \propto \rho^{1.6}, \\
& \omega_G \propto \sqrt{\frac{G_\text{cluster}}{\rho_\text{cluster}}} q_G = \sqrt{\frac{G_\text{cluster}}{\rho_\text{cluster}}} \frac{2\pi}{\xi_G}, \label{eq:frequencies}
\end{aligned}
\end{equation}
where $q_s = 2\pi/\xi_s$ is the characteristic wavenumber corresponding to $\xi_s$ and $\sqrt{G_0/\rho}$ represents the lowest value of the transverse sound speed in the gel, which is most relevant in the low-frequency regime.
$\rho_\text{cluster}$ and $G_\text{cluster}$ denote the density and shear modulus, respectively, of the glassy clusters, and thus, $\sqrt{{G_\text{cluster}}/{\rho_\text{cluster}}}$ represents the characteristic sound speed in the clusters.
As expressed in Eq.~(\ref{eq:frequencies}) and demonstrated in Fig.~\ref{fig_frequency}(a), $\omega_\ast$ follows a power-law scaling with the density, while $\omega_G$ is rather insensitive to the density.

Based on these two frequencies, a gel exhibits multiscale characteristics in its vibrational properties: (i) phonon-like vibrations in the homogeneous medium at low frequencies~($\omega < \omega_\ast$), (ii) phonon-like vibrations in the heterogeneous medium at intermediate frequencies~($\omega_\ast < \omega < \omega_G$), and (iii) disordered vibrations in the amorphously structured medium at high frequencies~($\omega > \omega_G$).
In particular, the phonon-like vibrations at $\omega < \omega_G$ exhibit crossover behavior in the dispersion curve at a frequency of $\omega_\ast$ and a wavelength of $\xi_s$ as follows:
\begin{equation}
\omega
\left\{ \begin{aligned}
&\propto \xi^{-1} & (\omega < \omega_\ast \ \&\ \xi > \xi_s),\\
&\propto \xi^{-a} & (\omega_\ast < \omega < \omega_G \ \&\ \xi_G < \xi < \xi_s),
\end{aligned} \right.~\label{eq:dispersion}
\end{equation}
where $a $~($> 1$) is the exponent of the dispersion curve of the phonon-like vibrations in the fractal structure~\cite{Alexander_1989,Yakubo_1989,percolation,Nakayama_1994}.

Accordingly, the vDOS $g(\omega)$ depends on $\omega$ differently between $\omega < \omega_\ast$ and $\omega > \omega_\ast$~(see Fig.~\ref{fig_frequency}(b)):
\begin{equation}
g(\omega)
\left\{ \begin{aligned}
&= A_D \omega^{2}~(=A_D \omega^{d-1}) & (\omega < \omega_\ast),\\
&\propto \omega^{\tilde{d}-1} & (\omega_\ast < \omega < \omega_G),
\end{aligned} \right.~\label{eq:specdim}
\end{equation}
where $A_D$ is the Debye level and $\tilde{d}$ is the number of spectral dimensions, which is defined as $\tilde{d}= D_f/a$, with $D_f$ being the number of fractal dimensions of the structure and $a$ being the exponent of the dispersion curve~\cite{Alexander_1989,Yakubo_1989,percolation,Nakayama_1994}.
The values of $\tilde{d}$, $D_f$, and $a$, all of which depend on $\rho$, have been reported in Table~I of Ref.~\onlinecite{Mizuno_2021}.
Finally, at $\omega > \omega_G$, the $\omega$ dependence of $g(\omega)$ is similar to that in a glass.

Here, we note that in Ref.~\onlinecite{Mizuno_2021}, we define $\omega_\ast$ and $\omega_G$ in a systematic way for both gels and glasses:
$\omega_\ast$ is defined as the boson-peak frequency or the frequency at which the vDOS converges to the Debye behavior, whereas $\omega_G$ is defined as the frequency at which the vibrational modes completely lose their phonon-like nature, with a zero value of the phonon order parameter.
Thus, $\omega_\ast$ and $\omega_G$ can also be defined for glasses.
However, their values are of the same order of magnitude, $\omega_\ast \sim \omega_G$, and are located near the boson-peak frequency~\cite{Mizuno_2021}~(see also filled symbols in Fig.~\ref{fig_frequency}(a)).
Glasses behave as elastic media with defects in the low-frequency regime below $\omega_\ast \sim \omega_G$~\cite{Lerner_2016,Mizuno_2017,Shimada_2018,Wang_2019}, while they behave as amorphously structured media in the high-frequency regime above $\omega_\ast \sim \omega_G$~\cite{Silbert_2009,Mizuno_2017,Shimada_2018,Wyart2_2005,Wyart_2006,Wyart_2005}.
Thus, the boson peak~\cite{Buchenau_1984,Yamamuro_1996,Mizuno2_2013,Mori_2020} at approximately $\omega_\ast \sim \omega_G$ points to the boundary between these two behaviors of glasses.

\section{Methods}~\label{sect:methods}
\subsection{System description}~\label{subsect:system}
The present work continues to study the system that has been studied in our previous paper~\cite{Mizuno_2021}.
The system is composed of $N$ point particles in three-dimensional~($d=3$) space under periodic boundary conditions in all three directions.
Particles $i$ and $j$ interact via the LJ potential:
\begin{align}
\phi_\text{LJ}(r) = 4\epsilon \left[ \left( \frac{\sigma_{ij}}{r} \right)^{12} - \left( \frac{\sigma_{ij}}{r} \right)^{6} \right], \label{eq:ljpotential}
\end{align}
where $r$ is the distance between the two particles and $\sigma_{ij} = (\sigma_i + \sigma_j)/2$, with $\sigma_i$ and $\sigma_j$ being the sizes~(diameters) of particles $i$ and $j$, respectively.
To avoid crystallization, we introduce polydispersity in the distribution of particle sizes~\cite{Leonforte2005}.
Specifically, the values of $\sigma_i$~($i=1,2,...,N$) are uniformly distributed in a range of $0.8 \sigma$ to $1.2 \sigma$.
The potential is cut off at $r = r_c = 3 \sigma$, where the potential and its first derivative are both made continuous as follows~\cite{Shimada_2018}:
\begin{align}
\phi(r) = \phi_\text{LJ}(r) - \phi_\text{LJ}(r_c) - (r-r_c) \frac{d\phi_\text{LJ}(r_c)}{dr}.
\end{align}
The mass $m$ is identical for all particles.
In the present paper, energy, length, and time are measured in units of $\epsilon$, $\sigma$, and $\tau = \sqrt{(m\sigma^2)/\epsilon}$, respectively.
Temperature and frequency are measured in units of $\epsilon/k_B$~($k_B$ is Boltzmann's constant) and $\tau^{-1} = \sqrt{\epsilon/(m\sigma^2)}$, respectively.

We vary the number density $\rho = N/V$ (where $V=L^3$ is the volume of the system and $L$ is the linear dimension) to study both gel and glass states.
The packing fraction $\varphi$ of the present polydisperse system is calculated as
\begin{align}
\varphi = \left( \frac{\pi}{6} \rho \right) \left( \int_{0.8\sigma}^{1.2\sigma} \frac{\sigma_i^3}{0.4\sigma} d\sigma_i \right). \label{eq:volumef}
\end{align}
The values of $\rho$~($\varphi$) are set to $1.0$~($0.54$), $0.7$~($0.38$), $0.5$~($0.27$), and $0.3$~($0.16$).
Additionally, we employ $N = 320000$ and $640000$ as the number of particles.

We first equilibrate the system in the normal liquid state at a temperature of $T=3.0$.
We then quench the system to the zero-temperature state of $T=0$ by minimizing the system potential and bringing the system to a local potential minimum.
Here, we employ the steepest descent method~\cite{Press_2007} for minimization.
We numerically consider the system to have settled to a local potential minimum when the maximum value $f_\text{max}$ among the forces $\left| \mathbf{F}_i \right|$ acting on all particles $i$~($i=1,2,...,N$) falls below $10^{-9}$.
Note that this protocol corresponds to an instantaneous quenching process with an infinite quenching rate.

In the following, we denote the $T=0$ configuration of the particles~(or the inherent structure) by $\mathbf{r} =\left[ \mathbf{r}_{1},\mathbf{r}_{2},...,\mathbf{r}_{N} \right]$~(a $3N$-dimensional vector), where $\mathbf{r}_{i}$ is the position of particle~$i$.
As demonstrated in Ref.~\onlinecite{Mizuno_2021}, we obtain the glass configuration at $\rho~(\varphi) = 1.0~(0.54)$ and the gel configurations at $\rho~(\varphi) = 0.7~(0.38)$, $0.5~(0.27)$, and $0.3~(0.16)$, where the gel configurations are generated through the arrested phase separation process~\cite{Zaccarelli2007,Lu2008,Zaccarelli_2008,Lu2013}.
Figure~\ref{fig_visual} presents the particle configuration of the gel state with $\rho = 0.3$~($\varphi = 0.16$) in (a) and that of the glass state with $\rho = 1.0$~($\varphi = 0.54$) in (b).

\subsection{Phonon transport analysis}
We perform phonon transport analysis in the same manner employed previously for studies of glasses~\cite{Gelin_2016,Mizuno_2018,Moriel_2019,Wang2_2019} and jammed particulate packings~\cite{Saitoh_2021}.
In this analysis, we simulate vibrational dynamics around the $T=0$ configuration, $\mathbf{r} =\left[ \mathbf{r}_{1},\mathbf{r}_{2},...,\mathbf{r}_{N} \right]$, in the harmonic approximation limit.
Below, we briefly explain this analysis~(please refer to Ref.~\onlinecite{Compute_vib} for details).

We denote by $\mathbf{u}_i(t)$ the displacement of particle $i$ from $\mathbf{r}_i$ at time $t$.
We first excite a phonon at the initial time $t=0$ by perturbing the velocity $\dot{\mathbf{u}}_i$ of particle $i$~(the dot denotes the time derivative) as follows:
\begin{equation}
\dot{\mathbf{u}}_i(t=0) = \dot{\mathbf{u}}^0_i = {\mathbf{a}}_\alpha \sin \left( \mathbf{q} \cdot \mathbf{r}_i + \psi \right),
\end{equation}
where $\mathbf{q}$ is the wave vector; $q \equiv \left| \mathbf{q} \right|$ is the wavenumber; $\alpha$ denotes the polarization, with $\alpha = L$ indicating longitudinal waves and $\alpha = T$ indicating transverse waves; and $\psi$ is set to $0$ or $\pi/2$.
The polarization vector (unit vector) ${\mathbf{a}}_\alpha$ is determined as $\mathbf{a}_L = {\mathbf{q}}/q$ for the longitudinal case and $\mathbf{a}_T \cdot {\mathbf{q}} = 0$ for the transverse case.

We next solve the linearized equation of motion:
\begin{equation}
\ddot{\mathbf{u}}_i = \sum_{j=1}^{N} \mathbf{D}_{ij} \cdot \mathbf{u}_j + \dot{\mathbf{u}}^0_i \delta (t),
\end{equation}
where $\mathbf{D}_{ij}$ is the dynamical matrix~\cite{Leibfried,Compute_vib} and $\delta(t)$ is the Dirac delta function.
From the time history of $\mathbf{u}_i(t)$, we calculate the normalized velocity--velocity correlation function:
\begin{equation}~\label{ctfunction0}
C(q,t) \equiv \frac{ {\left( \sum_{i=1}^{N} \dot{\mathbf{u}}_i(t) \cdot \dot{\mathbf{u}}_i^0 \right)} }{ {\left( \sum_{i=1}^{N} \dot{\mathbf{u}}^0_i \cdot \dot{\mathbf{u}}^0_i \right)} }.
\end{equation}
The function $C(q,t)$ represents the propagation and attenuation behaviors of the initially excited phonon $\dot{\mathbf{u}}^0_i$.

We performed repeated simulations of a phonon $\dot{\mathbf{u}}^0_i$ at a fixed wavenumber $q$ and polarization $\alpha$ by changing the wavevector to $\mathbf{q}=(q,0,0)$, $(0,q,0)$, and $(0,0,q)$ and using $\psi=0$ and $\pi/2$.
There are two independent transverse waves with different polarization vectors, $\mathbf{a}_{T_1}$ and $\mathbf{a}_{T_2}$~($\mathbf{a}_{T_1} \cdot \mathbf{a}_{T_2} = 0 $), which were also simulated independently.
The final value of $C(q,t)$ for a given $q$ and $\alpha$ was obtained by averaging over these simulation cases.
Note that since we implemented periodic boundary conditions in all three directions, $q$ takes discrete values of $q=(2\pi/L)n$, where $n=1,2,3,4,...$ is an integer.

In addition to the time correlation function $C(q,t)$, we also examine the Fourier transform of $C(q,t)$:
\begin{equation}~\label{ftcort1}
\tilde{C}(q,\omega) = \int_0^\infty C(q,t) \cos(\omega t) dt,
\end{equation}
where $\omega$ is the frequency.
We note that $\tilde{C}(q,\omega)$ corresponds to the so-called dynamical structure factor~\cite{Compute_vib}.
In general, $\tilde{C}(q,\omega)$ takes a maximum value at a certain frequency~(see Fig.~\ref{fig_spectrum}), and we define the average frequency $\bar{\omega}$ as
\begin{equation}~\label{ftcort2}
\bar{\omega} = \frac{ \int_{\omega_1}^{\omega_2} \omega  \tilde{C}(q,\omega) d\omega }{\int_{\omega_1}^{\omega_2} \tilde{C}(q,\omega) d\omega }\ (= \bar{\omega}(q)),
\end{equation}
where $\omega_1$ and $\omega_2$ are the lower and upper frequencies, respectively, at which $\tilde{C}(q,\omega)$ takes half its maximum value.
$\bar{\omega}=\bar{\omega}(q)$ gives an ``effective'' dispersion curve of phonon transport.

\subsubsection{Damped harmonic oscillator model~(DHOM)}
\begin{figure}[t]
\centering
\includegraphics[width=0.475\textwidth]{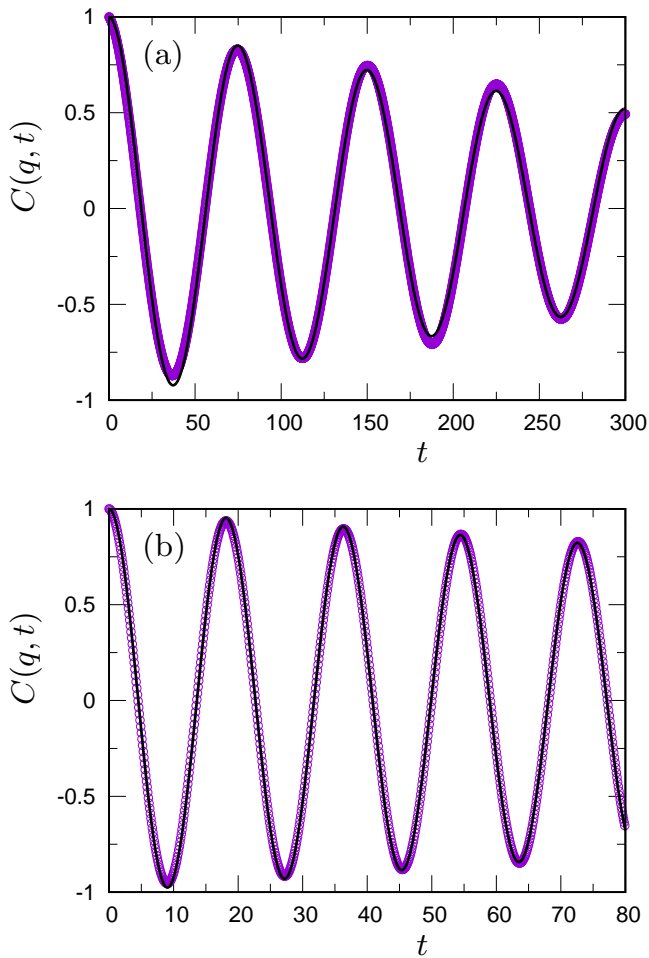}
\caption{\label{fig_ct}
{Time evolution of the velocity--velocity correlation function.}
Plots of $C(q,t)$ as a function of $t$ for transverse waves and low frequencies below $\omega_\ast$.
(a) $\rho=0.3$~(gel), $q= 6.1 \times 10^{-2}$, $\Omega = 8.4 \times 10^{-2}$, $\Gamma_T = 4.3 \times 10^{-3}$.
(b) $\rho=1.0$~(glass), $q= 9.2 \times 10^{-2}$, $\Omega = 3.5 \times 10^{-1}$, $\Gamma_T = 5.4 \times 10^{-3}$.
The symbols represent simulation data.
To quantify the frequency $\Omega$ and the attenuation rate $\Gamma_T$, the simulation data were fitted with the functional form $C(q,t) = \cos(\Omega t) e^{-\Gamma_T t/2}$; the fitting results are shown as solid lines.
}
\end{figure}

For the low-frequency regime, which ranges up to the same order of magnitude as the IR limit $\omega_{\alpha \text{IR}} \sim \omega_\ast$, the time evolution of $C(q,t)$ is well described by the damped harmonic oscillator model~(DHOM)~\cite{Gelin_2016,Mizuno_2018,Moriel_2019,Wang2_2019,Saitoh_2021}.
Specifically, we can use the functional form
\begin{equation}~\label{ctfunction}
C(q,t) = \cos(\Omega t) e^{-\Gamma_\alpha t/2}
\end{equation}
to fit simulation data of $C(q,t)$~(see Fig.~\ref{fig_ct}).
Alternatively, we can equivalently use the functional form
\begin{equation}~\label{ctfunction2}
\begin{aligned}
\tilde{C}(q,\omega) &= \int_0^\infty \cos(\Omega t) e^{-\Gamma_\alpha t/2} \cos( \omega t) dt,\\
&\approx  \frac{ \Gamma_\alpha \Omega^2 }{ (\omega^2-\Omega^2)^2 + \omega^2\Gamma^2_\alpha }
\end{aligned}
\end{equation}
to fit the data of $\tilde{C}(q,\omega)$.
In Eq.~(\ref{ctfunction2}), $\Omega \approx \omega$ and $\Gamma_\alpha \ll \Omega$ are assumed.
This fitting procedure quantifies the propagation frequency $\Omega$, the sound speed $c_\alpha = \Omega/q$, and the attenuation rate $\Gamma_\alpha$.
Note that the IR limit proves to approximately coincide with $\omega_\ast$, i.e., $\omega_{\alpha \text{IR}} \simeq \omega_\ast$~(see Figs.~\ref{fig_ir} and~\ref{fig_ir2}).
We also note that these values of $\Omega$, $c_\alpha$, and $\Gamma_\alpha$ are functions of the wavenumber $q$; alternatively, we can treat $c_\alpha$ and $\Gamma_\alpha$ as functions of $\Omega$ by transforming $q$ into $\Omega$ via the relation $\Omega = \Omega(q)$.
In the low-frequency~(low-wavenumber) regime, $c_\alpha$ should coincide with the value of $\bar{\omega}/q$ obtained through Eq.~(\ref{ftcort2}).

\subsubsection{Ioffe--Regel (IR) limit}
\begin{figure}[t]
\centering
\includegraphics[width=0.475\textwidth]{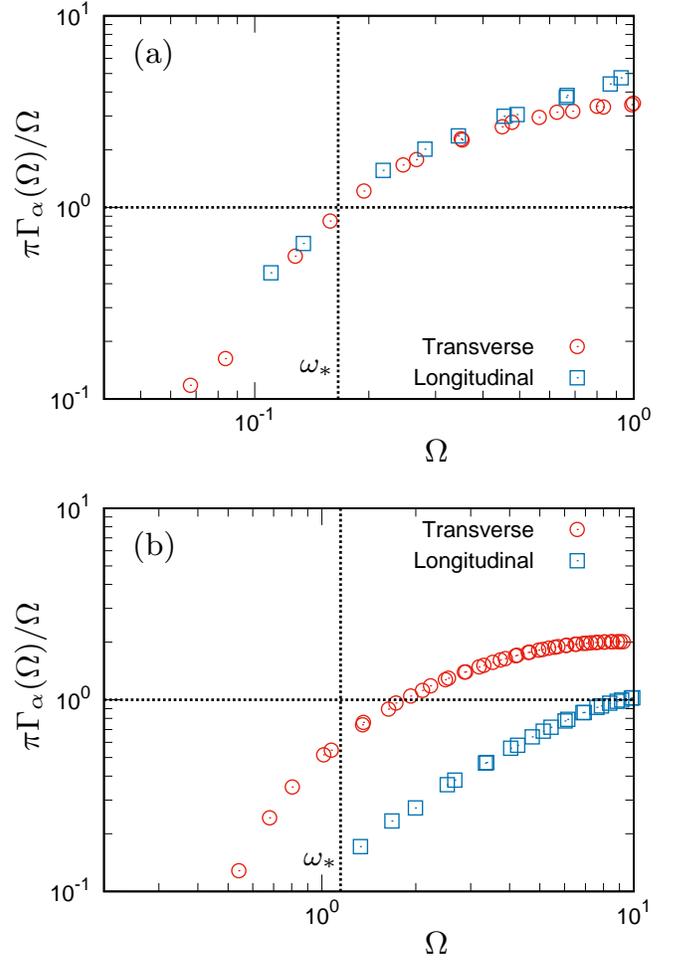}
\caption{\label{fig_ir}
{The IR limit.}
The ratio $\pi \Gamma_\alpha(\Omega)/\Omega$ is plotted as a function of $\Omega$ for transverse waves ($\alpha =T$, red circles) and longitudinal waves ($\alpha = L$, blue squares).
(a) $\rho = 0.3$~(gel).
(b) $\rho = 1.0$~(glass).
The frequency at which this ratio is equal to one (i.e., $\pi \Gamma_\alpha(\Omega)/\Omega = 1$) is defined as the IR frequency, $\omega_{\alpha \text{IR}}$.
In the gel (a), we observe $\omega_{T\text{IR}} = \omega_{L\text{IR}}$, which can thus be uniformly denoted by $\omega_\text{IR}$ by dropping the index $\alpha$.
$\omega_\ast$ coincides with $\omega_\text{IR}$~($= \omega_{T\text{IR}} = \omega_{L\text{IR}}$) in the gel (a), whereas in the glass (b), $\omega_\ast \simeq \omega_{T\text{IR}}$ but $\omega_\ast \ll \omega_{L\text{IR}}$.
}
\end{figure}

\begin{figure}[t]
\centering
\includegraphics[width=0.475\textwidth]{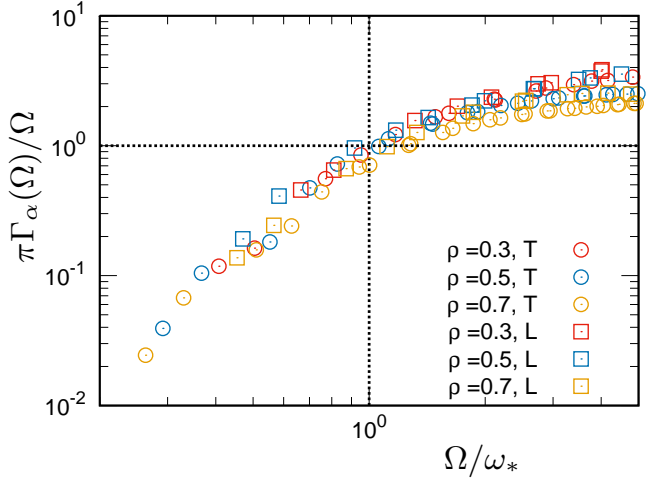}
\caption{\label{fig_ir2}
{The IR limit for gels with different densities.}
The ratio $\pi \Gamma_\alpha(\Omega)/\Omega$ is plotted as a function of $\Omega/\omega_\ast$ for transverse waves ($\alpha =T$, circles) and longitudinal waves ($\alpha = L$, squares) and for different densities of $\rho =0.3$, $0.5$, and $0.7$.
We observe that $\omega_{\text{IR}}/\omega_\ast=\omega_{T\text{IR}}/\omega_\ast = \omega_{L\text{IR}}/\omega_\ast \simeq 1$, i.e., $\omega_\text{IR} = \omega_{T\text{IR}} = \omega_{L\text{IR}} \simeq \omega_\ast$, regardless of the density value.
}
\end{figure}

Here, we define an important frequency scale, the IR limit $\omega_{\alpha \text{IR}}$~($\alpha = T$ or $L$), as~(see Figs.~\ref{fig_ir} and~\ref{fig_ir2})
\begin{equation}
\frac{\pi \Gamma_\alpha(\Omega = \omega_{\alpha \text{IR}})}{ \omega_{\alpha \text{IR}}} = 1,
\end{equation}
where $\Gamma_\alpha (\Omega)$ is considered a function of $\Omega$.
Above $\Omega = \omega_{\alpha \text{IR}}$, the phonon decay time ($= \Gamma_\alpha^{-1}$) becomes shorter than half of the vibrational period ($= \pi/\Omega$); i.e., the phonon decays within half of the duration of one period.
The IR frequency $\omega_{\alpha \text{IR}}$ therefore corresponds to an upper bound on the propagation frequency of a phonon as a plane wave.
For frequencies up to the same order of magnitude as $\omega_{\alpha \text{IR}}$, the DHOM can work; however, it does not necessarily work at high frequencies far above $\omega_{\alpha \text{IR}}$~\cite{Damart_2017,Beltukov_2018}.

\begin{figure}[t]
\centering
\includegraphics[width=0.475\textwidth]{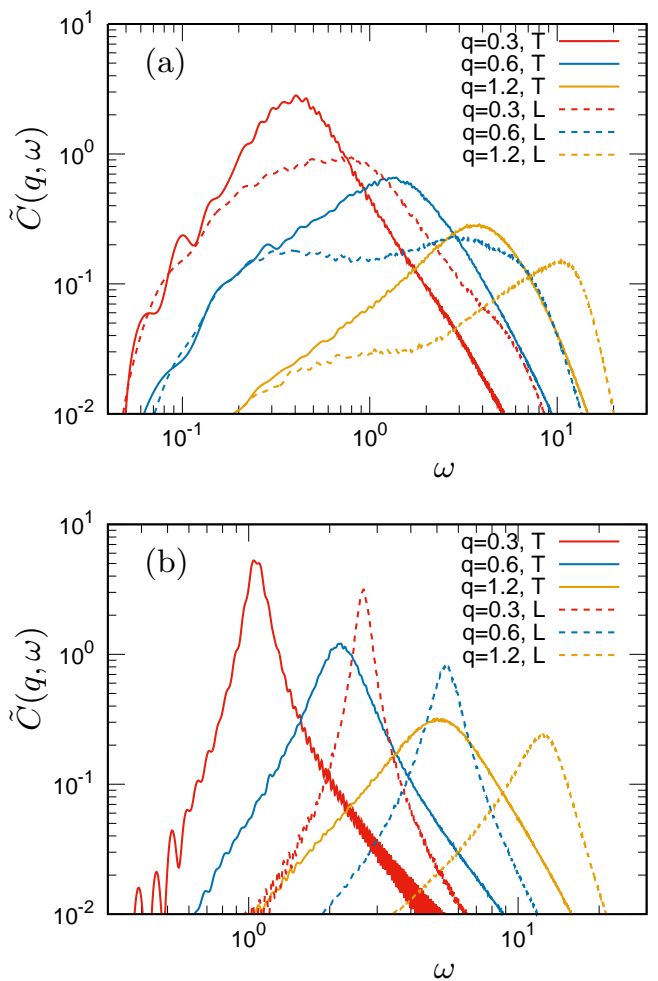}
\caption{\label{fig_spectrum}
{Fourier transform of the velocity--velocity correlation function.}
Plots of $\tilde{C}(q,\omega)$ as a function of $\omega$ for transverse~($T$) and longitudinal~($L$) waves and the indicated values of $q$.
(a) $\rho =0.3$~(gel) and (b) $\rho=1.0$~(glass).
}
\end{figure}

\begin{figure*}[t]
\centering
\includegraphics[width=0.995\textwidth]{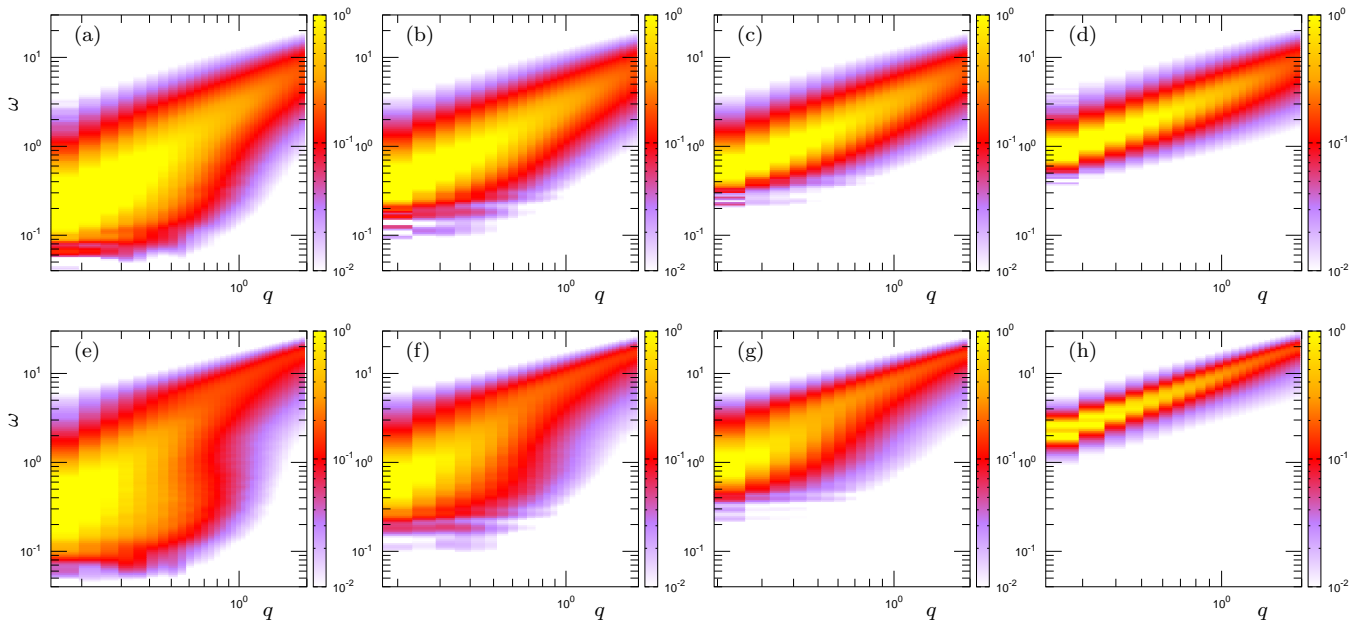}
\caption{\label{fig_spectrum2}
{Contour plots of the Fourier transform of the velocity--velocity correlation function.}
Plots of $\tilde{C}(q,\omega)$ as a function of $q$ and $\omega$ for transverse waves~(upper panels, (a) to (d)) and longitudinal waves~(lower panels, (e) to (h)).
The density is $\rho = 0.3$ in (a, e), $0.5$ in (b, f), $0.7$ in (c, g), and $1.0$~(glass) in (d, h).
}
\end{figure*}

\begin{figure}[t]
\centering
\includegraphics[width=0.475\textwidth]{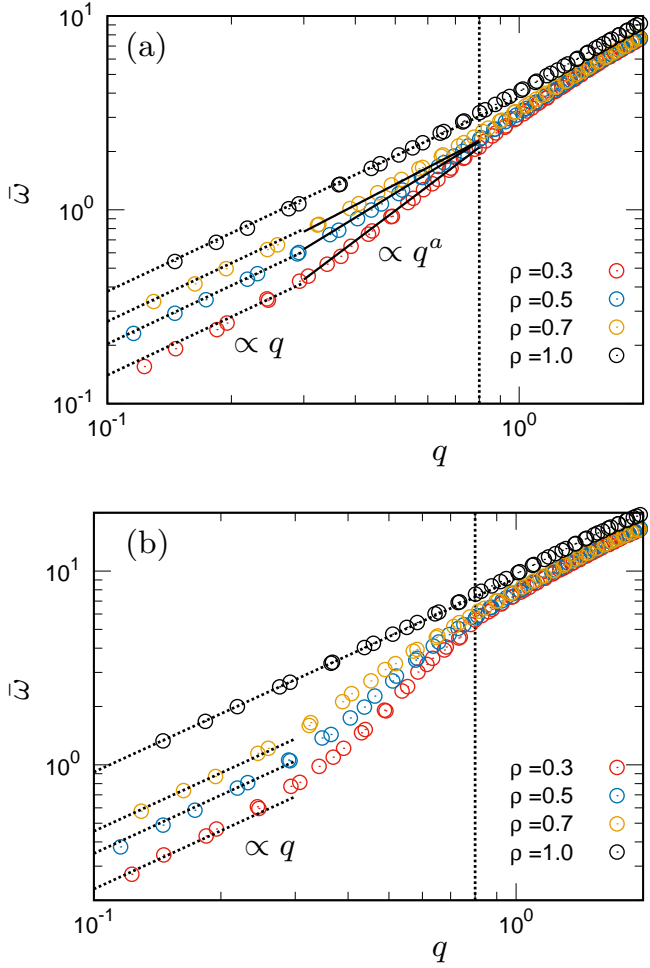}
\caption{\label{fig_effsp}
{Effective dispersion curves.}
Plots of $\bar{\omega}$ as a function of $q$ for (a) transverse waves~($\alpha =T$) and (b) longitudinal waves~($\alpha =L$) and for different densities $\rho =0.3$, $0.5$, $0.7$~(gels) and $1.0$~(glass).
The vertical line indicates $q = 0.8 \simeq q_G$, which corresponds to the size of the glassy clusters, $\xi_G \simeq 7.8$~($q_G = 2\pi/\xi_G$).
The dotted lines indicate the linear dispersion curves of $\bar{\omega}=c_{\alpha 0} q \propto q$, where $c_{\alpha 0}$ is the macroscopic sound speed given in Eq.~(\ref{macrospeed}), while the solid lines in (a) indicate the dispersion curves of $\bar{\omega} \propto q^a$ with $a=1.1$, $1.3$, and $1.6$ for $\rho = 0.7$, $0.5$, and $0.3$, respectively.
These dispersion curves are the same as those of Eq.~(\ref{eq:dispersion}).
}
\end{figure}

\begin{figure}[t]
\centering
\includegraphics[width=0.475\textwidth]{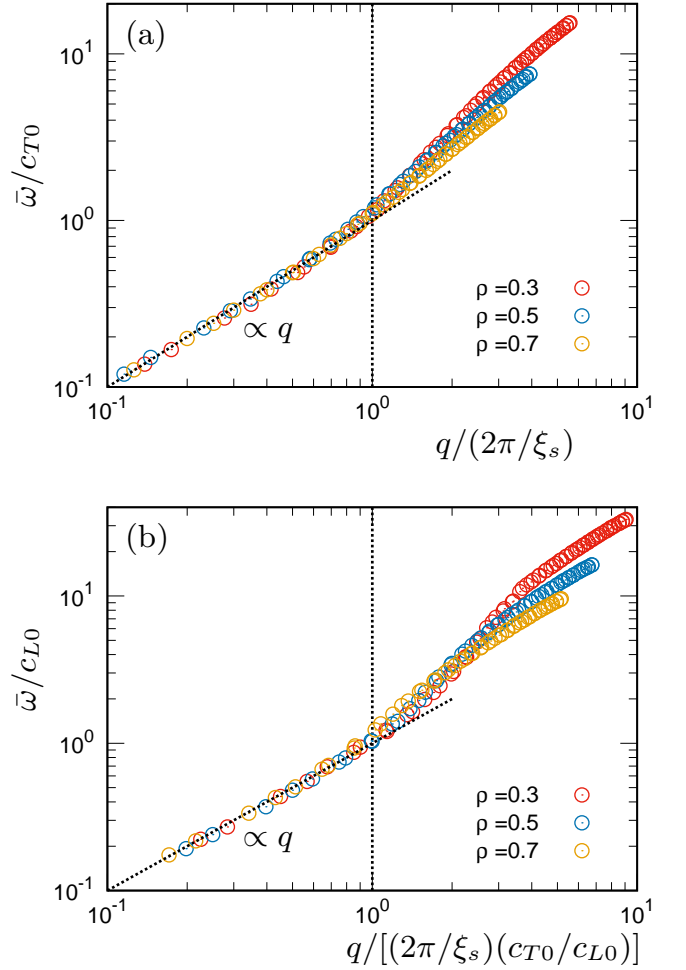}
\caption{\label{fig_effsp2}
{Scaled effective dispersion curves of gels.}
Plots of $\bar{\omega}/c_{\alpha 0}$ as a function of (a) $q/(2\pi/\xi_s)$ for transverse waves~($\alpha =T$) and (b) $q/[(2\pi/\xi_s)(c_{T0}/c_{L0})]$ for longitudinal waves~($\alpha =L$), with the indicated values of $\rho =0.3$, $0.5$, and $0.7$.
The data are the same as those in Fig.~\ref{fig_effsp}, but here, they are scaled using the values of $c_{\alpha 0}$ and $\xi_s$.
The vertical line indicates $q/(2\pi/\xi_s) = 1$ in (a) and $q/[(2\pi/\xi_s)(c_{T0}/c_{L0})]=1$ in (b).
We observe that the dispersion curves start to deviate from the initial linear curve~(dotted line) at approximately $q = 2\pi/\xi_s$ in (a) and $q = (2\pi/\xi_s)(c_{T0}/c_{L0})$ in (b).
}
\end{figure}

\begin{figure*}[t]
\centering
\includegraphics[width=0.98\textwidth]{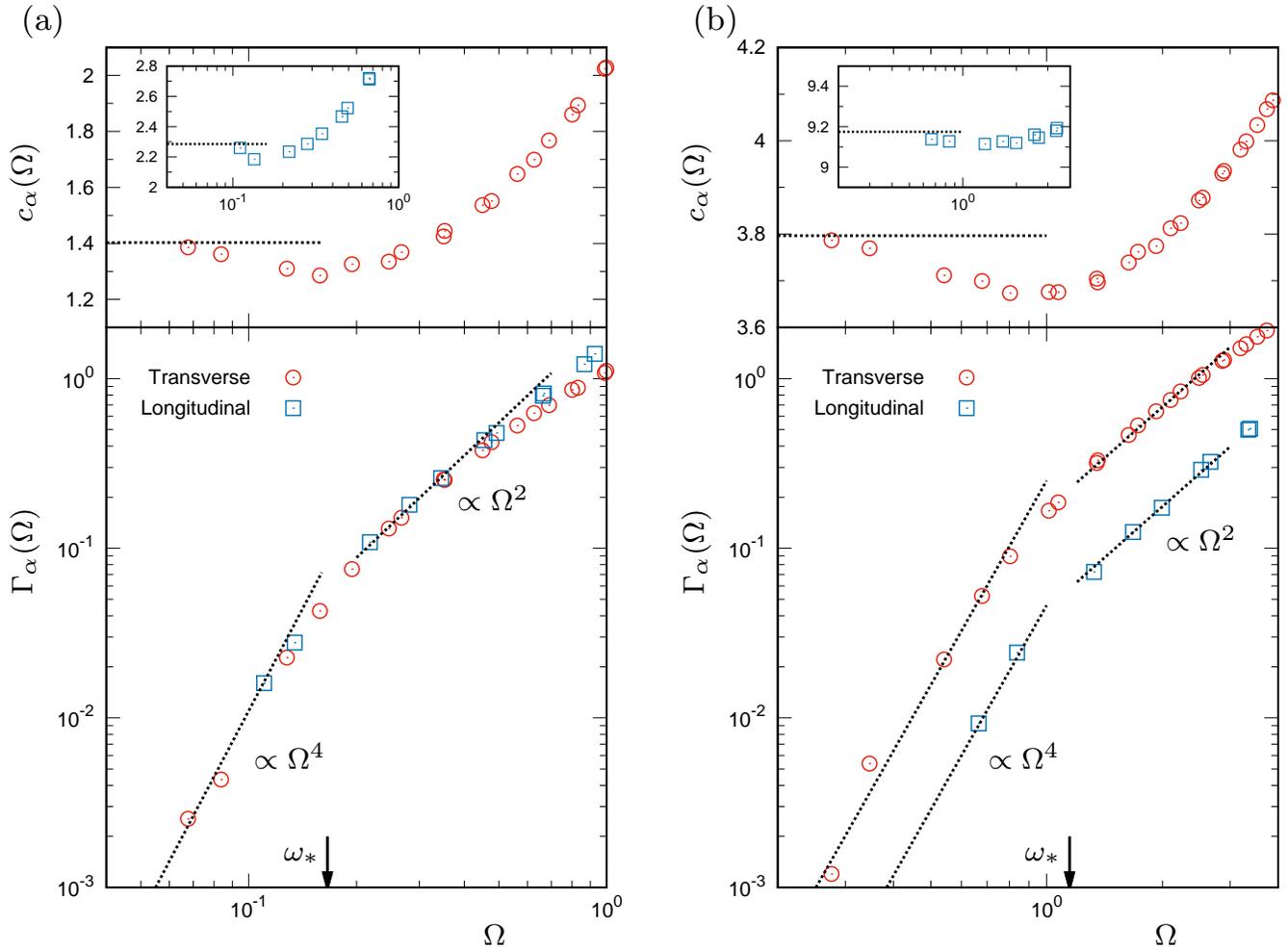}
\caption{\label{fig_speed}
{Sound speed and attenuation rate.}
Plots of $c_\alpha (\Omega)$ and $\Gamma_\alpha (\Omega)$ as functions of $\Omega$ for transverse waves ($\alpha =T$, red circles) and longitudinal waves ($\alpha = L$, blue squares).
(a, left panels) $\rho = 0.3$~(gel).
(b, right panels) $\rho = 1.0$~(glass).
In the lower panels, the frequency $\omega_\ast$ is indicated by an arrow.
The dotted lines indicate $c_\alpha (\Omega) = c_{\alpha 0}$ in the upper panels and the scalings of $\Gamma_\alpha \propto \Omega^4$ and $\propto \Omega^2$ in the lower panels.
}
\end{figure*}

\begin{figure}[t]
\centering
\includegraphics[width=0.475\textwidth]{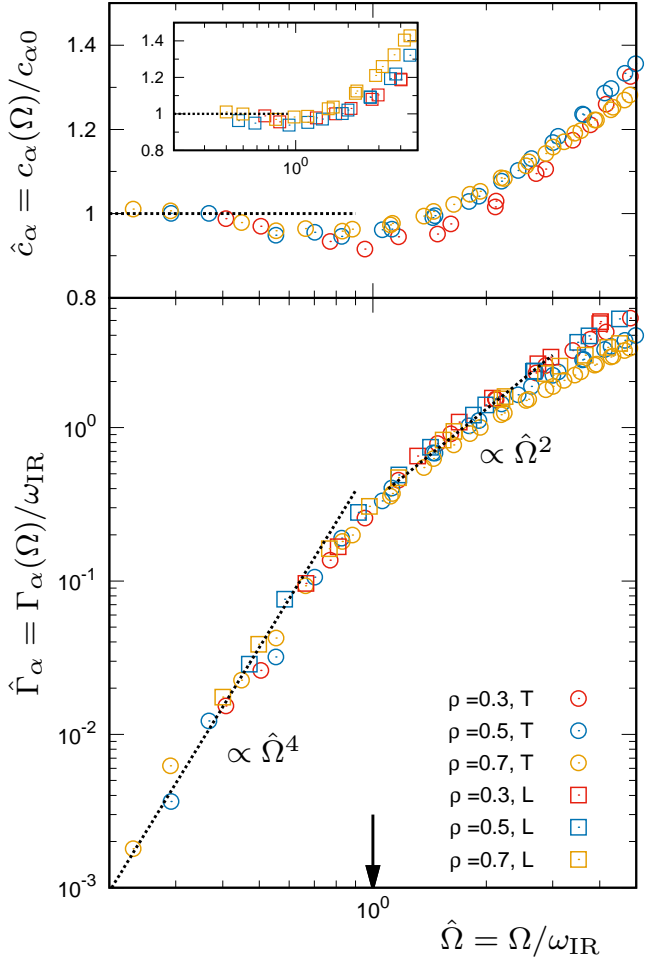}
\caption{\label{fig_speed2}
{Scaled sound speed and attenuation rate in gels.}
Plots of $\hat{c}_\alpha = c_\alpha(\Omega)/c_{\alpha 0}$ and $\hat{\Gamma}_\alpha = \Gamma_\alpha(\Omega)/\omega_\text{IR}$ as functions of $\hat{\Omega} = \Omega/\omega_\text{IR}$ for transverse waves ($\alpha =T$, circles) and longitudinal waves ($\alpha = L$, squares) and for different densities of $\rho =0.3$, $0.5$, and $0.7$.
In the lower panel, the arrow indicates $\hat{\Omega}=1$, i.e., $\Omega = \omega_\text{IR}$.
The dotted lines indicate $\hat{c}_\alpha =1$ in the upper panel and the scalings of $\hat{\Gamma}_\alpha \propto \hat{\Omega}^4$ and $\propto \hat{\Omega}^2$ in the lower panel.
In the low-frequency regime, the data converge well for different densities.
Note that $\omega_\text{IR}$~($= \omega_{T\text{IR}} = \omega_{L\text{IR}}$) is nearly equal to $\omega_\ast$, $\omega_\text{IR} \simeq \omega_\ast$.
}
\end{figure}

\begin{figure}[t]
\centering
\includegraphics[width=0.475\textwidth]{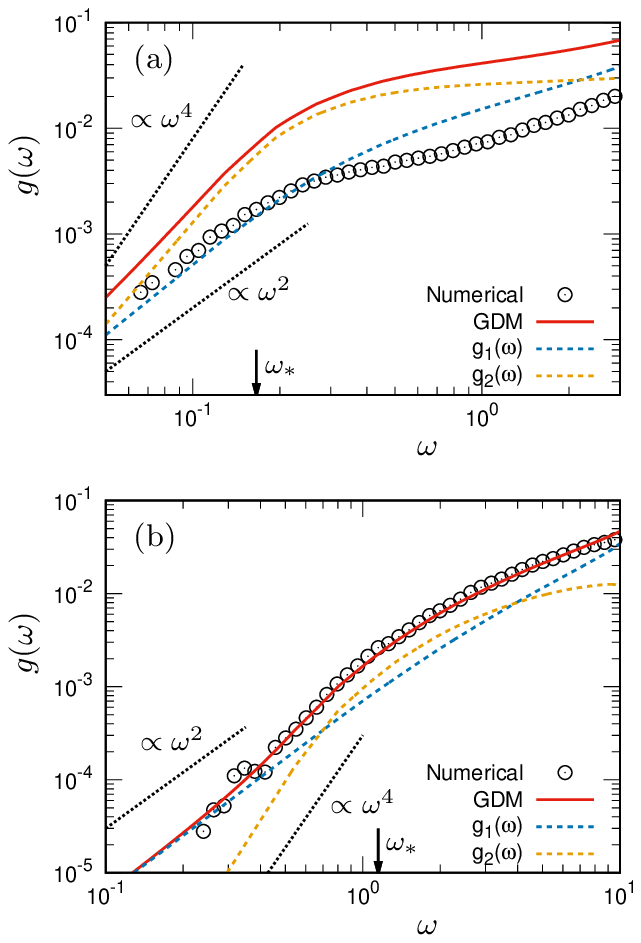}
\caption{\label{fig_rdf}
{Comparison between the numerical vDOS values obtained through vibrational mode analysis and the vDOS of the GDM.}
Plots of $g(\omega)$ as a function of $\omega$.
(a) $\rho = 0.3$~(gel).
(b) $\rho = 1.0$~(glass).
The symbols represent the numerical values from the vibrational mode analysis reported in Ref.~\onlinecite{Mizuno_2021}.
The lines represent the vDOS of the GDM as given in Eq.~(\ref{pvdos}).
The vDOS in Eq.~(\ref{pvdos}) is composed of two terms, one related to the dispersion curve ($g_1(\omega) = f_1 \left[ \omega; c_\alpha(\omega) \right]$) and the other related to sound broadening ($g_2(\omega) = f_2 \left[ \omega; c_\alpha(\omega),\Gamma_\alpha(\omega) \right]$).
The GDM correctly predicts the numerical values in the glass (b), but in the gel (a), it greatly overestimates the values due to the term $g_2(\omega)$.
The frequency $\omega_\ast$ is indicated by an arrow.
}
\end{figure}

\section{Results}
\subsection{Correlation function}
Figure~\ref{fig_ct} shows examples of data on the time evolution of the correlation function $C(q,t)$~[Eq.~(\ref{ctfunction0})] for the gel with $\rho =0.3$ in (a) and the glass with $\rho = 1.0$ in (b).
The data in this figure are in the low-frequency regime below $\omega_\ast$, where the DHOM does apply.
As shown in the figure, we can fit the functional form of Eq.~(\ref{ctfunction})~(solid line) to the simulation data~(symbols) for both the gel and glass states.
This fitting procedure quantifies the propagation frequency $\Omega$, the sound speed $c_\alpha = \Omega/q$, and the attenuation rate $\Gamma_\alpha$, which are presented in Figs.~\ref{fig_speed} and~\ref{fig_speed2} and will be discussed in Sec.~\ref{subsec:lowfreq}.

We also present examples of data on the Fourier transform of $C(q,t)$, $\tilde{C}(q,\omega)$~[Eq.~(\ref{ftcort1})], in Fig.~\ref{fig_spectrum}.
$\tilde{C}(q,\omega)$ shows a sharper peak in its $\omega$ dependence as the wavenumber $q$ decreases.
For sufficiently low values of $q$, the DHOM function in Eq.~(\ref{ctfunction2}) should work, and $\Omega$ and $\Gamma_\alpha$ correspond to the location of the peak and its full width at half maximum, respectively.
However, for large $q$, $\tilde{C}(q,\omega)$ shows a rather broad spectrum, and the DHOM does not necessarily work.

Figure~\ref{fig_spectrum2} shows $\tilde{C}(q,\omega)$ for the entire $q$ and $\omega$ regime, except for the lowest $q$ and $\omega$ values~(where the numerical precision of the Fourier transform is not good).
We can clearly recognize that the spectra systematically broaden as the density becomes lower in the gel states~($\rho = 0.7$ to $0.3$).
In a comparison between the data for the gel with $\rho=0.3$ in (a,e) and the glass with $\rho=1.0$ in (d,h), the gel shows much broader spectra than the glass at a given $q$ and polarization $\alpha$~($= T$ or $L$)~(compare also the data between panels~(a) and (b) of Fig.~\ref{fig_spectrum}).
This observation indicates that phonons are much more strongly scattered in lower-density gels than in the dense glass state.
This can be attributed to the porous structures of gels~(see Fig.~\ref{fig_visual}), i.e., phonons are strongly scattered at the holes throughout their structures.

When looking at the gels, e.g., the gel with $\rho = 0.3$ in panels (a,e) of Fig.~\ref{fig_spectrum2}, in more detail, we notice that at some high $q \sim 1$, the spectra abruptly become narrow and approach those of the glass state with $\rho =1.0$.
As visualized in Fig.~\ref{fig_visual}(a), clusters of glasses become connected to form a network structure in the gel state.
At $q \sim 1$, the wavelength becomes comparable to the size of these glassy clusters, i.e., $q \sim q_G \simeq 0.8$, and phonons of $q \gtrsim q_G$ enter these clusters.
This behavior of the phonons is reflected by the abrupt narrowing of $\tilde{C}(q,\omega)$ at $q \sim 1$.
In this high-$q$ regime, we observe that the effective dispersion curves of $\bar{\omega}(q)$ rapidly become similar between the gels and the glass, as shown in Figs.~\ref{fig_effsp} and~\ref{fig_effsp2} and discussed in Sec.~\ref{subsec:highfreq} below.

\subsection{Effective dispersion curves}~\label{subsec:highfreq}
Figure~\ref{fig_effsp} presents the data on the effective dispersion curves of $\bar{\omega}=\bar{\omega}(q)$~[Eq.~(\ref{ftcort2})] for both the gels and the glass and for transverse waves~($\alpha =T$) in (a) and longitudinal waves~($\alpha =L$) in (b).
Note that even for cases in which phonons \textit{cannot} propagate as plane waves, we can define the effective dispersion curve of $\bar{\omega} = \bar{\omega}(q)$ as in Eq.~(\ref{ftcort2}).
From this figure, we can confirm the linearity of the dispersion curve, $\bar{\omega} = c_{\alpha 0}q \propto q$, in the low-$q$ regime, where $c_{\alpha 0}$~($\alpha =T, L$) is the macroscopic sound speed as
\begin{equation} \label{macrospeed}
c_{T0}=\sqrt{\frac{G}{\rho}}, \qquad
c_{L0} = \sqrt{\frac{(K+4G/3)}{\rho}},
\end{equation}
where we employ the average value of the shear modulus, i.e., $G = G_\text{ave}$, since we perform the phonon transport analysis by averaging over the directions of the wavevector $\mathbf{q}$.
In the glass with $\rho =1.0$, this linear curve persists up to a high value of $q \sim 1$.
In contrast, for the gels, as $q$ increases, the dispersion curve starts to deviate from the linear curve at a certain wavenumber $q = q_{\alpha \ast}$.
For transverse waves~[Panel (a)] at higher values of $q \gtrsim q_{T \ast}$, we present the dispersion curves, $\bar{\omega} \propto q^a$, with $a=1.1$, $1.3$, and $1.6$ for $\rho = 0.7$, $0.5$, and $0.3$, respectively.
These dispersion curves of $\bar{\omega} = c_{\alpha 0}q$ at lower $q$ and $\bar{\omega} \propto q^a$ at higher $q$ are consistent with those given in Eq.~(\ref{eq:dispersion})
\footnote{
The values of $a$ coincide between those extracted from vibrational eigenmodes~\cite{Mizuno_2021} and those extracted from transverse waves.
This is because the low-frequency eigenmodes exhibit a more transverse than longitudinal nature due to the small shear modulus values.
}.

Figure~\ref{fig_effsp2} plots the scaled value of $\bar{\omega}$ versus the scaled wavenumber, i.e., $\bar{\omega}/c_{T0}$ versus $q/(2\pi \xi_s)$ for transverse waves in Panel (a) and $\bar{\omega}/c_{L0}$ versus $q/[(2\pi \xi_s)(c_{T0}/c_{L0})]$ for longitudinal waves in (b).
Note that the factor of $c_{T0}/c_{L0}$ in the scaled $q$ for longitudinal waves is independent of $\rho$ and thus gives only a constant offset, since $c_{T0}$ and $c_{L0}$ follow the same scaling with $\rho$ because both the bulk and shear moduli follow the same scaling with $\rho$, namely, $K \propto G \propto \rho^{2.5}$.
We recognize that the scaled $q$ at which the dispersion curve starts to deviate from linearity is located at approximately $1$ regardless of the density value.
This observation yields $q_{T \ast} = 2\pi/\xi_s$ for transverse waves and $q_{L \ast} = (2\pi/\xi_s)(c_{T0}/c_{L0})$ for longitudinal waves.
$q_{T \ast}$ and $q_{L \ast}$ both follow the same power-law scaling with $\rho$,
\begin{equation}
q_{T \ast} \propto q_{L \ast} \propto \xi_s^{-1} \propto \rho^{0.7},
\end{equation}
and we therefore conclude that $q_{T \ast}$ and $q_{L \ast}$ both correspond to the same length scale $\xi_s$, which is the boundary between the homogeneous medium and the heterogeneous medium with a fractal structure.

In Fig.~\ref{fig_effsp}, as $q$ increases further and reaches approximately $q = 0.8$, $\bar{\omega}$ shows a crossover to approach that of the glass with $\rho = 1.0$ for both transverse and longitudinal waves.
As we have reported in our previous work~\cite{Mizuno_2021} and described in Sec.~\ref{sec:preliminaries}, in gels, glassy clusters are dispersed in space, and these clusters are connected to form a network structure.
The value of $q = 0.8 \simeq q_G$ corresponds to the size of these clusters, which is insensitive to the density $\rho$.
We therefore conclude that phonons of $q > q_G \simeq 0.8$ pass into the glassy clusters and behave as though they are traveling in a glass.
At $q > q_G \simeq 0.8$, the dispersion curve is approximately linear and can be described as $\bar{\omega} = c_{\alpha \text{cluster}}q$, where $c_{\alpha \text{cluster}}$ is the sound speed in the glassy clusters:
\begin{equation} \label{speedcluster}
\begin{aligned}
c_{T \text{cluster}} &= \sqrt{\frac{G_\text{cluster}}{\rho_\text{cluster}}}, \\
c_{L \text{cluster}} &= \sqrt{\frac{(K_\text{cluster}+4G_\text{cluster}/3)}{\rho_\text{cluster}}}.
\end{aligned}
\end{equation}
Here, $\rho_\text{cluster}$ is the density of the glassy clusters, and $G_\text{cluster}$ and $K_\text{cluster}$ are the shear and bulk moduli of the clusters, respectively~(see also Eq.~(\ref{eq:frequencies})).
Thus, from another point of view, we can measure the elastic moduli of the clusters from phonon transport at high wavenumbers of $q > q_G \simeq 0.8$.
Note that the $\bar{\omega}$ of a gel takes a lower value than the corresponding glass value, which means that the elastic moduli of the glassy clusters are lower than that of bulk glass with $\rho =1.0$.

In summary, gels show a multiscale nature in their effective dispersion curves: (i) the dispersion curve is linear, $\bar{\omega} = c_{\alpha 0}q$, at low wavenumbers~($q < q_{\alpha \ast}$); (ii) the dispersion curve is fractal-like, $\bar{\omega} \propto q^a$, at intermediate wavenumbers~($q_{\alpha \ast} < q < q_G$); and (iii) the dispersion curve in the glassy clusters predominates at high wavenumbers~($q > q_G$).
This multiscale property is controlled by the two length scales $\xi_s$ and $\xi_G$ in the static structure, as $q_{\alpha \ast} \propto \xi_s^{-1}$ and $q_{G} \propto \xi_G^{-1}$.
In turn, as in Eq.~(\ref{eq:frequencies}), the frequency $\omega_\ast$ is associated with $q_{\alpha \ast}$, whereas $\omega_G$ is associated with $q_G$.

\subsection{Sound speed and attenuation rate}~\label{subsec:lowfreq}
Beginning in this section, we focus on the low-frequency~(low-wavenumber) regime, in which phonons can propagate as plane waves and the DHOM does apply.
More specifically, the DHOM works in a frequency regime that extends up to the same order of magnitude as the IR limit $\omega_{\alpha \text{IR}}$.
As will be shown in Figs.~\ref{fig_ir} and~\ref{fig_ir2} and next Sec.~\ref{subsec:ir}, the IR limit of a gel takes the same value for both transverse and longitudinal waves and is approximately the same as $\omega_\ast$, i.e., $\omega_{\text{IR}} =\omega_{T \text{IR}} = \omega_{L \text{IR}} \simeq \omega_\ast$.
We thus focus on the frequency regime up to approximately $\omega_\ast$, which corresponds to the wavenumber regime of $q \lesssim q_{\alpha \ast}$ where the linear dispersion curve of $\bar{\omega} = c_{\alpha 0}q \propto q$ persists, as shown in Figs.~\ref{fig_effsp} and~\ref{fig_effsp2}.

Figure~\ref{fig_speed} shows the sound speed $c_\alpha = \Omega/q$~(upper panels) and the attenuation rate $\Gamma_\alpha$~(lower panels) as functions of the propagation frequency $\Omega$.
For the glass with $\rho=1.0$, as seen in the right panels, the sound speed for transverse waves shows a clear softening at approximately $\omega_\ast$~(i.e., the boson-peak frequency~\cite{Mizuno_2021}).
Additionally, the attenuation data show a crossover at approximately $\omega_\ast$, from Rayleigh scattering behavior ($\Gamma_\alpha \propto \Omega^4$) to diffusive damping behavior ($\Gamma_\alpha \propto \Omega^2$).
These observations are fully consistent with previously reported simulations~\cite{Monaco2_2009,Marruzzo_2013,Mizuno_2014,Mizuno_2018}.
We note that the emergence of softening in longitudinal waves depends on the details of the system~\cite{Mizuno_2014} and is not observed in the present system.

Turning our attention to the gel with $\rho =0.3$ in the left panels of Fig.~\ref{fig_speed}, we observe that transverse and longitudinal waves behave very similarly.
The sound speeds for both types of waves show a clear softening at approximately $\omega_\ast$, whereas the attenuation rate shows a crossover at $\omega_\ast$ from Rayleigh scattering ($\Gamma_\alpha \propto \Omega^4$) to diffusive damping ($\Gamma_\alpha \propto \Omega^2$).
Remarkably, both transverse and longitudinal waves show identical values of $\Gamma_\alpha$, in marked contrast to the case in glasses, where transverse attenuation is much stronger than longitudinal attenuation.
This scattering behavior in gel states can be attributed to their porous structure~(see Fig.~\ref{fig_visual}(a)), i.e., both transverse and longitudinal waves are strongly scattered at the holes throughout a gel.

To analyze phonon transport in gels of different densities, we introduce scaled quantities of the frequency, sound speed, and attenuation rate:
\begin{equation}
\begin{aligned}
& \hat{\Omega} = \frac{\Omega}{\omega_{\text{IR}}} \simeq \frac{\Omega}{\omega_\ast}, \\
& \hat{c}_\alpha = \frac{c_\alpha}{c_{\alpha 0}}, \\
& \hat{\Gamma}_\alpha = \frac{\Gamma_\alpha}{\omega_{\text{IR}}} \simeq \frac{\Gamma_\alpha}{\omega_\ast},
\end{aligned}
\end{equation}
where we recall that $\omega_{\text{IR}} =\omega_{T \text{IR}} = \omega_{L \text{IR}} \simeq \omega_\ast$.
Figure~\ref{fig_speed2} plots $\hat{c}_\alpha$ and $\hat{\Gamma}_\alpha$ as functions of $\hat{\Omega}$.
We observe that the data for different densities nearly converge to a single curve, except for the high-frequency data far above $\omega_\ast$.
We thus establish
\begin{equation} \label{eq:lowerfreq}
\begin{aligned}
&c_{T,L}(\Omega) \approx c_{T,L 0}, \\
&\Gamma_{T,L}(\Omega) \approx 
\left\{ \begin{aligned}
& A \omega_\ast \left( \frac{\Omega}{\omega_\ast} \right)^4 & (\Omega \lesssim \omega_\ast), \\
& B \omega_\ast \left( \frac{\Omega}{\omega_\ast} \right)^2 & (\Omega \gtrsim \omega_\ast),
\end{aligned} \right. 
\end{aligned}
\end{equation}
where $A \simeq 0.59$ and $B \simeq 0.33$ are constants that do \textit{not} depend on the polarization $\alpha = T$ or $L$.
These results demonstrate that the density dependencies of the sound speeds are determined~(trivially) by the elastic moduli, whereas those of scattering are controlled by the characteristic frequency $\omega_\ast \simeq \omega_{\text{IR}}$~(which is associated with the length scale $\xi_s$ of the static structure), regardless of the transverse or longitudinal nature of the waves.

\subsection{Ioffe--Regel~(IR) limit}~\label{subsec:ir}
We next discuss the IR frequency $\omega_{\alpha \text{IR}}$ in gels.
Figure~\ref{fig_ir} presents $\pi \Gamma_\alpha(\Omega)/\Omega$ as a function of $\Omega$ for the gel with $\rho =0.3$ in (a) and the glass with $\rho = 1.0$ in (b).
We recall that at $\Omega = \omega_{\alpha \text{IR}}$, $\pi \Gamma_\alpha(\Omega)/\Omega$ crosses $1$.
For the glass in Panel (b), the $\omega_{T\text{IR}}$ value for transverse waves coincides with the boson-peak frequency, $\omega_{T\text{IR}} \approx \omega_\ast$, while the $\omega_{L\text{IR}}$ value for longitudinal waves is located at a much higher frequency, $\omega_{L\text{IR}} \gg \omega_{T\text{IR}} \approx \omega_\ast$.
These results have been reported in many previous papers on glasses\cite{Mizuno_2018,Monaco2_2009,Marruzzo_2013,Mizuno_2014,Wang_2015,Beltukov_2016}.
We note that longitudinal waves can propagate in a glass even far above the boson peak.

Turning to the gel with $\rho = 0.3$, it is remarkable that both transverse and longitudinal waves point to an identical value of the IR limit frequency, which coincides well with $\omega_\ast$, i.e., $\omega_{T \text{IR}} = \omega_{L \text{IR}} \simeq \omega_\ast$.
Figure~\ref{fig_ir2} also presents $\pi \Gamma_\alpha(\Omega)/\Omega$ as a function of $\Omega/\omega_\ast$ for gels of different densities.
The data for transverse and longitudinal waves as well as different densities collapse to a single curve, again yielding $\omega_{T \text{IR}} = \omega_{L \text{IR}} \simeq \omega_\ast$.
We therefore simply denote the IR limit for both transverse and longitudinal waves by $\omega_{\text{IR}}$~(dropping the index $\alpha$), and thus, $\omega_{\text{IR}} = \omega_{T\text{IR}} = \omega_{L\text{IR}} \simeq \omega_\ast$.

As described in Sec.~\ref{sec:preliminaries}, the vibrational modes are phonon-like both above and below $\omega_\ast$~\cite{Mizuno_2021}, i.e., below $\omega_\ast$, the phonon vibrations are associated with the homogeneous medium, while above $\omega_\ast$, they are associated with the heterogeneous medium with a fractal structure.
It is remarkable that even above the IR limit $\omega_\ast \simeq \omega_{\text{IR}}$, a phonon-like nature persists in the vibrational modes.
Nevertheless, phonons above $\omega_\ast$, which are composed of such phonon-like modes, actually \textit{cannot} propagate as plane waves.
This peculiar result can be attributed to scattering at the holes in the porous structure; i.e., even though the vibrational forms are phonon-like, they instantaneously attenuate at the holes.
We thus conclude that phonon localizations of both transverse and longitudinal waves emerge at frequencies above $\omega_\ast$~(or wavelengths below $\xi_s$) in gel states.

\subsection{Generalized Debye model~(GDM)}~\label{subsec:tgdm}
Here, we discuss the validity of the GDM for gels.
The GDM can predict the vDOS by using data on the sound speed $c_\alpha(\Omega)$ and the attenuation rate $\Gamma_\alpha(\Omega)$~\cite{schirmacher_2006,schirmacher_2007,schirmacher_2015,Wyart_2010,DeGiuli_2014}.
In short, the GDM assumes the phonon approximation $\Omega \gg \Gamma_\alpha$ and formulates the Green function $\tilde{G}_\alpha(q,\omega)$ in terms of $c_\alpha(\omega)$ and $\Gamma_\alpha(\omega)$ as shown in Eq.~(\ref{greenf3}); then, it provides the vDOS $g(\omega)$ as given in Eq.~(\ref{3dpvdos}):
\begin{equation} \label{pvdos}
\begin{aligned}
g(\omega) & = g_1(\omega) + g_2(\omega), \\
& \equiv f_1 \left[ \omega; c_\alpha(\omega) \right] + f_2 \left[ \omega; c_\alpha(\omega),\Gamma_\alpha(\omega) \right],
\end{aligned}
\end{equation}
where $g_1(\omega) = f_1 \left[ \omega; c_\alpha(\omega) \right]$ is a functional of $c_\alpha(\omega)$, whereas $g_2(\omega) = f_2 \left[ \omega; c_\alpha(\omega),\Gamma_\alpha(\omega) \right]$ is a functional of both $c_\alpha(\omega)$ and $\Gamma_\alpha(\omega)$.
Please refer to Appendix~\ref{sec:gdm} and Eq.~(\ref{3dpvdos}) for the details of the formulations.

The GDM has been shown to work well for glasses~\cite{Marruzzo_2013,Mizuno_2018}.
The simulation data in Figs.~\ref{fig_speed} and~\ref{fig_speed2} show that the phonon approximation $\Omega \gg \Gamma_\alpha$, which is a central assumption of the GDM, is valid for both the gels and the glass, at least in the low-frequency region of $\Omega \lesssim \omega_\ast$.
We thus may expect that the GDM can also work for the gels.
Figure~\ref{fig_rdf} compares the data from simulations~(symbols) with the predictions of the GDM~(red solid lines) for the gel with $\rho = 0.3$ in (a) and the glass with $\rho = 1.0$ in (b).
As we have already reported in our previous work~\cite{Mizuno_2018} and as is confirmed in Panel (b), the GDM can quantitatively capture the vDOS of a glass.

In contrast to the glass case, we see that the GDM does \textit{not} work at all for the gel, as shown in Panel (a) of Fig.~\ref{fig_rdf}.
The GDM~(red solid line) predicts values much larger than the simulation data~(circles).
Instead, the Debye prediction~(blue dashed line), which does not include any contribution from attenuation, captures the simulation data at low frequencies of $\omega < \omega_\ast$.
At $\omega < \omega_\ast$, the vibrational modes are phonon-like in the homogeneous medium, so it is reasonable that the Debye prediction can capture the vDOS, as mentioned in our previous work~\cite{Mizuno_2021}.
We therefore conclude that the inconsistency between the GDM and the simulation data arises from the attenuation contribution.

We consider a possible reason why the GDM works for glasses but not for gels, as follows.
The GDM estimates the contribution to the vDOS from the attenuation rate~($g_2(\omega)$), which is added to the Debye value~($g_1(\omega)$).
This contribution~($g_2(\omega)$) originates from the fact that attenuation arises when a phonon is composed of additional eigenmodes with different frequencies, i.e., a finite value of the attenuation rate means an increase in the value of the vDOS.
For the gel case, phonons are sharply cut off at the interfaces of holes.
To produce this cut-off behavior for phonons, we need a complicated superposition of additional eigenmodes in different frequency regimes.
This effect leads to strong phonon scattering and causes the GDM to overestimate the vDOS.
We thus consider that the porous structure of gels causes strong attenuation, which then leads to overestimation by the GDM.

\section{Discussion and conclusions}~\label{sect:conclusion}
As a complement to our previous paper~\cite{Mizuno_2021}, the present work has studied the phonon transport properties of particulate physical gels (LJ gels).
We have demonstrated that the multiscale characteristics of the static structure and vibrational states extend to phonon transport.
The phonon transport behavior shows two distinct crossovers at frequencies of $\omega_\ast$ and $\omega_G$, which correspond to wavenumbers of $q_{\alpha \ast} \sim \xi_s^{-1}$ and $q_G \sim \xi_G^{-1}$, respectively.
This behavior is characterized by (i) a linear dispersion curve and Rayleigh scattering at $\omega < \omega_\ast$ or $q < q_{\alpha \ast}$, (ii) the dispersion curve of the fractal structure and diffusive damping at $\omega_\ast < \omega < \omega_G$ or $q_{\alpha \ast} < q < q_G$, and (iii) phonon transport through the glassy clusters at $\omega > \omega_G$ or $q > q_G$.
The IR limit is located at the remarkably low frequency $\omega_\ast$.
Thus, phonon transport is localized above $\omega_\ast$ or $q_{\alpha \ast}$, and phonons in regimes (ii) and (iii) do \textit{not} propagate as plane waves; rather, their dynamics are diffusive.

The present results are fully consistent with experimental observations in silica aerogels~\cite{Vacher_1990,Courtens_1987,Courtens_1988,Vacher_1989,Anglaret_1994}.
Silica gels also exhibit two characteristic frequencies at which both the vDOS and phonon transport show crossovers.
In particular, their phonon transport is characterized by Rayleigh scattering in the low-frequency regime, and the IR limit frequency is located at the edge of the Rayleigh scattering regime.
The present work has established that these multiscale behaviors in the vDOS and phonon transport are controlled by static structural properties, as the two characteristic frequencies~($\omega_\ast$ and $\omega_G$) are associated with the two length scales~($\xi_s$ and $\xi_G$) in the structure.

One remarkable result is that phonon transport is localized when the phonons enter the fractal structure~(regime (ii), at $\omega_\ast < \omega < \omega_G$ or $q_{\alpha \ast} < q < q_G$).
Experimental works~\cite{Vacher_1990,Courtens_1987,Courtens_1988,Vacher_1989,Anglaret_1994} have explained this phonon localization in terms of fractons, which are highly localized vibrations in the fractal structure~\cite{Alexander_1989,Yakubo_1989,percolation,Nakayama_1994}.
Contrary to this explanation, our simulation results have demonstrated that the vibrational modes are extended~(\textit{not} localized)~\cite{Mizuno_2021}, thus differing in nature from fractons.
This observation is somewhat strange at first glance, since phonons are mainly composed of these extended vibrational states.
We consider this to be attributable to the porous structure of gels, as follows.
When a phonon propagates through this porous structure, its vibrational state is cut off sharply at the interfaces of holes.
To produce this cut-off behavior of a phonon, a complicated superposition of different vibrational states in different frequency regimes is needed.
Thus, the phonon is composed of additional vibrational modes with different frequencies, so it is strongly attenuated.
This mechanism of strong scattering~(due to the porous structure) also causes Rayleigh scattering to arise in regime (i) at $\omega < \omega_\ast$ or $q < q_{\alpha \ast}$, even where the vDOS follows the Debye law with phonon-like modes.
In addition, it causes breakdown of the GDM: the GDM does \textit{not} work for gels, although it does work for glasses.

Thus, in a gel, the scattering at the holes in the porous structure is dominant, and consequently, phonon transport is controlled by the static structure.
For either transverse or longitudinal waves, their behaviors are controlled by the same characteristic lengths of $\xi_s$ and $\xi_G$ in the structure.
This is in marked contrast to phonon transport in glasses, which is controlled by elastic heterogeneities~\cite{Marruzzo_2013,Mizuno_2014,schirmacher_2006,schirmacher_2007,schirmacher_2015}.
These elastic heterogeneities produce excess vibrational modes over what is predicted by the Debye law, such as the boson peak~\cite{Buchenau_1984,Yamamuro_1996,Mizuno2_2013,Mori_2020} and quasi-localized modes~\cite{Lerner_2016,Mizuno_2017,Shimada_2018,Wang_2019}, which then cause phonon scattering, i.e., the excess vDOS causes a phonon to be composed of more modes with different frequencies, leading to a broader line width in the spectra and scattering.
Since the spatial heterogeneities are different between the shear and bulk modulus distributions~\cite{Mizuno_2013,Mizuno_2016,Shakerpoor_2020}, the strength of scattering and the characteristic length scales differ between transverse and longitudinal waves in glasses~\cite{Mizuno_2018,Silbert_2005}.

For future works, the viscoelastic properties of gels, which have been measured in many previous experiments~\cite{Trappe2000,Prasad_2003,Koumakis_2011,Hsiao_2012,fernandez2016fluids,Rocklin_2021}, will be important research targets.
Since viscoelasticity is directly related to vibrational states~\cite{Lemaitre_2006} and phonon transport~\cite{simpleliquid,Mizuno_2012}, we expect that multiscale characteristics will also manifest in this phenomenon, which can then be understood based on the structural properties of gels, as demonstrated here for the phonon transport properties.

Another important point is that the present system is an aerogel, whose representative example is silica gel~\cite{Vacher_1990,Courtens_1987,Courtens_1988,Vacher_1989,Anglaret_1994}; it can also be regarded as a porous glass~\cite{Phani_1987,Kovaci_2001,Niyogi_2021}.
In contrast, solvating media usually exist in colloidal gels, which have been an active topic in experimental and theoretical research.
Thus, it is important to consider particulate systems with solvating media and investigate hydrodynamic effects on the properties of gels.
Several works have studied hydrodynamic effects on the gelation of colloidal systems~\cite{Yamamoto_2008,Furukawa_2010,Cao_2012,Tateno2021}.
In particular, Refs.~\onlinecite{Furukawa_2010} and~\onlinecite{Cao_2012} reported that hydrodynamic interactions between colloidal particles lower the volume fraction threshold for percolation.
Ref.~\onlinecite{Tateno2021} pointed out that slow solvent transport~(permeation of the solvent) plays a crucial role in the colloidal gelation process.
In addition, and interestingly, Ref.~\onlinecite{Varga_2018} studied hydrodynamic effects on the vibrational states of colloidal gels.
The authors found that long-range hydrodynamic interactions make the eigenmodes more collective and extended.
Now, several computer simulation tools have been developed for particulate systems with solvating media~\cite{Tanaka_2000,Nakayama_2005,Kim_2006}.
Although the computational cost~(which is generally heavy) must be addressed, these tools can enable us to understand hydrodynamic effects from a microscopic point of view.

\section*{Acknowledgments}
This work was supported by JSPS KAKENHI Grant Numbers 18H05225, 19K14670, 19H01812, 20H01868, and 20H00128.

\section*{Author declarations}
The authors declare no conflicts of interest.

\section*{Data availability}
The data that support the findings of this study are available within the article.

\appendix

\begin{figure}[t]
\centering
\includegraphics[width=0.475\textwidth]{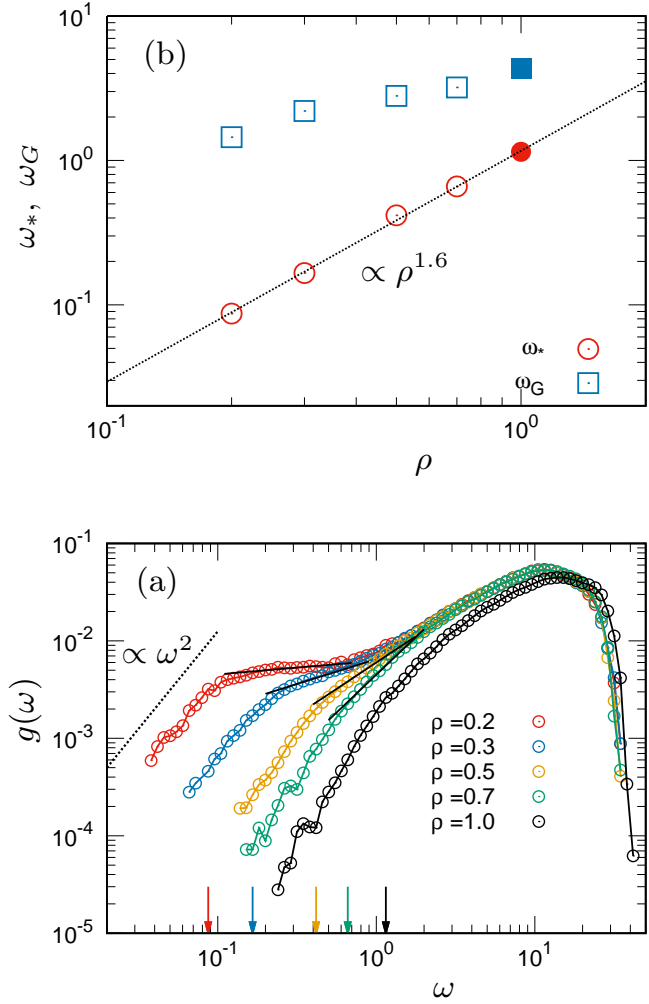}
\caption{\label{fig_frequency}
{Characteristic frequencies.}
(a) $\omega_\ast$ and $\omega_G$ are plotted as functions of $\rho$.
Filled symbols represent data for the glass with $\rho=1.0$.
The line represents the power-law scaling of $\omega_\ast \propto \rho^{1.6}$.
(b) The vDOS $g(\omega)$ is plotted as a function of $\omega$ for densities of $\rho = 0.2$, $0.3$, $0.5$, $0.7$~(gels) and $\rho = 1.0$~(glass).
Arrows indicate the corresponding values of $\omega_\ast$.
The solid black lines superimposed on the data for gels with $\rho = 0.2$ to $0.7$ represent $g(\omega) \propto \omega^{\tilde{d}-1}$ with the spectral dimension $\tilde{d}$.
}
\end{figure}

\section{Characteristic frequencies}~\label{sec:cfreq}
Here, we present data on the two characteristic frequencies, $\omega_\ast$ and $\omega_G$.
As described in Sec.~\ref{sec:preliminaries}, gels show multiscale characteristics in their vibrational states: (i) phonon-like vibrations in the homogeneous media at low frequencies~($\omega < \omega_\ast$), (ii) phonon-like vibrations in the heterogeneous media at intermediate frequencies~($\omega_\ast < \omega < \omega_G$), and (iii) disordered vibrations in the glassy clusters at high frequencies~($\omega > \omega_G$).
The frequencies $\omega_\ast$ and $\omega_G$ are the boundaries between (i) and (ii) and between (ii) and (iii), respectively.

Figure~\ref{fig_frequency}(a) plots the values of $\omega_\ast$ and $\omega_G$ as functions of the density $\rho$.
This figure demonstrates the power-law scaling of $\omega_\ast \propto \rho^{1.6}$ and that $\omega_G$ is insensitive to $\rho$, as described from Eq.~(\ref{eq:frequencies}).
We also present the vDOS $g(\omega)$ in Fig.~\ref{fig_frequency}(b).
$g(\omega)$ shows three distinct dependences on $\omega$: (i) the Debye behavior of $g(\omega) = A_D \omega^2$~($=A_D\omega^{d-1}$) at $\omega < \omega_\ast$, (ii) $g(\omega)\propto \omega^{\tilde{d}-1}$ with spectral dimension $\tilde{d}$ at $\omega_\ast < \omega < \omega_G$, and (iii) behavior similar to that in a glass at $\omega>\omega_G$.
We remark that $g(\omega)$ converges to the Debye vDOS $A_D \omega^2$ at $\omega < \omega_\ast$~\cite{Mizuno_2021}.
Additionally, $g(\omega)\propto \omega^{\tilde{d}-1}$ with $\tilde{d} = D_f/a$ works well at $\omega_\ast < \omega < \omega_G$~(see the solid black lines in Fig.~\ref{fig_frequency}(b)).

\section{Generalized Debye model~(GDM)}~\label{sec:gdm}
In this Appendix, we describe the GDM, the validity of which is tested for gels in Fig.~\ref{fig_rdf} and Sec.~\ref{subsec:tgdm}.
At low frequencies below $\omega_{\alpha \text{IR}}$, the vDOS can be formulated in the framework of the GDM~\cite{Marruzzo_2013,Mizuno_2018,schirmacher_2006,schirmacher_2007,schirmacher_2015,Wyart_2010,DeGiuli_2014}.
In the GDM, we assume the phonon approximation $\Omega \gg \Gamma_\alpha$, which has been shown to be valid in the low-frequency regime $\Omega \lesssim \omega_\ast$ for both gels and glasses, as seen in Figs.~\ref{fig_speed} and~\ref{fig_speed2}.

The Green function is defined as~\cite{Leibfried}
\begin{equation}
G(t) \equiv \frac{ \left( \sum_{i=1}^{N} {\mathbf{u}}_i(t) \cdot \dot{\mathbf{u}}_i^0 \right) }{ \left( \sum_{i=1}^{N} \dot{\mathbf{u}}^0_i \cdot \dot{\mathbf{u}}^0_i \right) } = \int C(q,t) dt.
\end{equation}
Using the functional form of $C(t)$ in Eq.~(\ref{ctfunction}), we obtain
\begin{equation}
G(t) \approx  \left\{ \frac{\sin(\Omega t)}{\Omega} \right\} e^{-\Gamma_\alpha t/2} H(t),
\end{equation}
where $H(t)$ is the Heaviside step function.
The Fourier transform of $G(t)$ is formulated as
\begin{equation} \label{greenf2}
\tilde{G}_\alpha(q,\omega) = \int_{-\infty}^{+\infty} G(t) e^{i\omega t} dt \approx \frac{1}{-\omega^2 + q^2 \hat{c}_\alpha(q,\omega)^2},
\end{equation}
where $\hat{c}_\alpha(q,\omega)$ is the complex sound speed:
\begin{equation}
\hat{c}_\alpha(q,\omega) = c_\alpha(q) \left\{ 1 - i \frac{\omega \Gamma_\alpha(q)}{\Omega(q)^2} \right\}^{1/2}.
\end{equation}
Then, following Ref.~\onlinecite{Marruzzo_2013}, we approximate $\hat{c}_\alpha(q,\omega)$ by dropping the dependence on the wavenumber $q$, as follows:
\begin{equation}
\begin{aligned}
\hat{c}_\alpha(q,\omega) &\approx \hat{c}_\alpha \left(q=\Omega^{-1}(\omega),\omega \right), \\
 &= c_\alpha(\omega) \left\{ 1 - i \frac{\Gamma_\alpha(\omega)}{\omega} \right\}^{1/2},
\end{aligned}
\end{equation}
where $\Omega^{-1}$ denotes the inverse function, $c_\alpha(\omega) = c_\alpha(q=\Omega^{-1}(\omega))$, and $\Gamma_\alpha(\omega) = \Gamma_\alpha(q=\Omega^{-1}(\omega))$.
Note that this approximation is valid within the phonon approximation, $\Omega(q) \gg \Gamma_\alpha (q)$.
We finally formulate $\tilde{G}_\alpha(q,\omega)$ as
\begin{equation} \label{greenf3}
\tilde{G}_\alpha(q,\omega) \approx \frac{1}{ \left\{ 1 - i \frac{\Gamma_\alpha(\omega)}{\omega} \right\} \left\{ -\omega^2 + q^2 c_\alpha(\omega )^2 \right\} }.
\end{equation}

By using the Green function $\tilde{G}_\alpha(q,\omega)$, the vDOS can be formulated as
\begin{equation}
\begin{aligned}
g(\omega) &= \left({2\omega}/{\pi q_D^3} \right) \\
& \ \times \int_{0}^{q_D} dq q^{2} \text{Im} \left\{ 2\tilde{G}_T({q},\omega) + \tilde{G}_L({q},\omega) \right\},
\end{aligned}
\end{equation}
where $q_D = \sqrt[3]{6\pi^{2} {\rho}}$ is the Debye wavenumber.
Using $\tilde{G}_\alpha(q,\omega)$ in Eq.~(\ref{greenf3}), we derive the following:
\begin{equation}
\begin{aligned} \label{3dpvdos}
& g(\omega) = \left( \frac{2}{c_T^3 q_D^3} + \frac{1}{c_L^3 q_D^3} \right) \omega^2 \\
& + \frac{2}{\pi} \Bigg[
\frac{2\Gamma_T}{c_T^2 q_D^2} \left\{ 1 + \left( \frac{\omega}{2c_T q_D}\right) \log\left( \frac{ c_T q_D-\omega}{c_T q_D+\omega} \right) \right\} \\
& +
\frac{\Gamma_L}{c_L^2 q_D^2} \left\{ 1 + \left( \frac{\omega}{2c_L q_D}\right) \log\left( \frac{ c_L q_D-\omega}{c_L q_D+\omega} \right) \right\}
\Bigg].
\end{aligned}
\end{equation}
The first and second terms in Eq.~(\ref{3dpvdos}) define $g_1(\omega) = f_1\left[ \omega; c_\alpha(\omega) \right]$ and $g_2(\omega) = f_2\left[\omega; c_\alpha(\omega),\Gamma_\alpha(\omega) \right]$, respectively, in Eq.~(\ref{pvdos}).
Note that the first term $g_1(\omega)$ corresponds to the phonon~(Debye) vDOS, whereas the second term $g_2(\omega)$ represents the additional, nonphonon vDOS that arises from the broadening~(attenuation).

\bibliographystyle{apsrev4-2}
\bibliography{reference}

\end{document}